\documentclass[superscriptaddress,aps,prl,twocolumn,preprintnumbers,amsmath,amssymb,floatfix,showpacs]{revtex4-1}

\usepackage{graphicx}% Include figure files
\usepackage{dcolumn}% Align table columns on decimal point
\usepackage{bm}% bold math
\usepackage{amssymb}
\usepackage{subfigure}
\usepackage{xr}

\newcommand{\ket}[1]{\ensuremath{\left|#1\right\rangle}}

\begin{document}

\title{Generation of squeezed Schr\"odinger cat states\\ with an operation allowing iterative growth}

\author{Jean Etesse}

\affiliation{Laboratoire Charles Fabry, Institut d'Optique, CNRS, Universit\'e Paris Sud \\ 2 avenue Augustin Fresnel, 91127 Palaiseau cedex, France}

\author{Martin Bouillard}

\affiliation{Laboratoire Charles Fabry, Institut d'Optique, CNRS, Universit\'e Paris Sud \\ 2 avenue Augustin Fresnel, 91127 Palaiseau cedex, France}

\author{Bhaskar Kanseri}

\affiliation{Laboratoire Charles Fabry, Institut d'Optique, CNRS, Universit\'e Paris Sud \\ 2 avenue Augustin Fresnel, 91127 Palaiseau cedex, France}

\author{Rosa Tualle-Brouri}

\affiliation{Laboratoire Charles Fabry, Institut d'Optique, CNRS, Universit\'e Paris Sud \\ 2 avenue Augustin Fresnel, 91127 Palaiseau cedex, France}
\affiliation{Institut Universitaire de France, 103 boulevard Saint-Michel, 75005 Paris, France}

\date{\today}

%%%%%%%%%%%%%%%%%%%%%%%%%%%%%%%%%%%%%%%

\begin{abstract}
We present what is to our knowledge the first implementation of a Schr\"odinger cat states ``breeding" operation, which allows an iterative growth of these states. We thus report the experimental generation of a squeezed Schr\"odinger cat state from two single photon Fock states, which can be seen as cat states with zero amplitude. These Fock states are mixed on a symmetrical beamsplitter and the generation is heralded by a homodyne measurement in one of the two output arms. The output state has a fidelity of 61\% with an even squeezed Schr\"odinger cat state of amplitude $\alpha=1.63$. This hybrid operation opens up new prospects in quantum optics as the protocol depicted here can be iterated in order to produce new kind of mesoscopic states.
\end{abstract}

%%%%%%%%%%%%%%%%%%%%%%%%%%%%%%%%%%%%%%%%

\maketitle
%%%%%%%%%%%%%%%%%%%%%%%%%%%%%%%%%%%%%%%%

\begin{figure}[!h]
\begin{center}
\includegraphics[width=6cm]{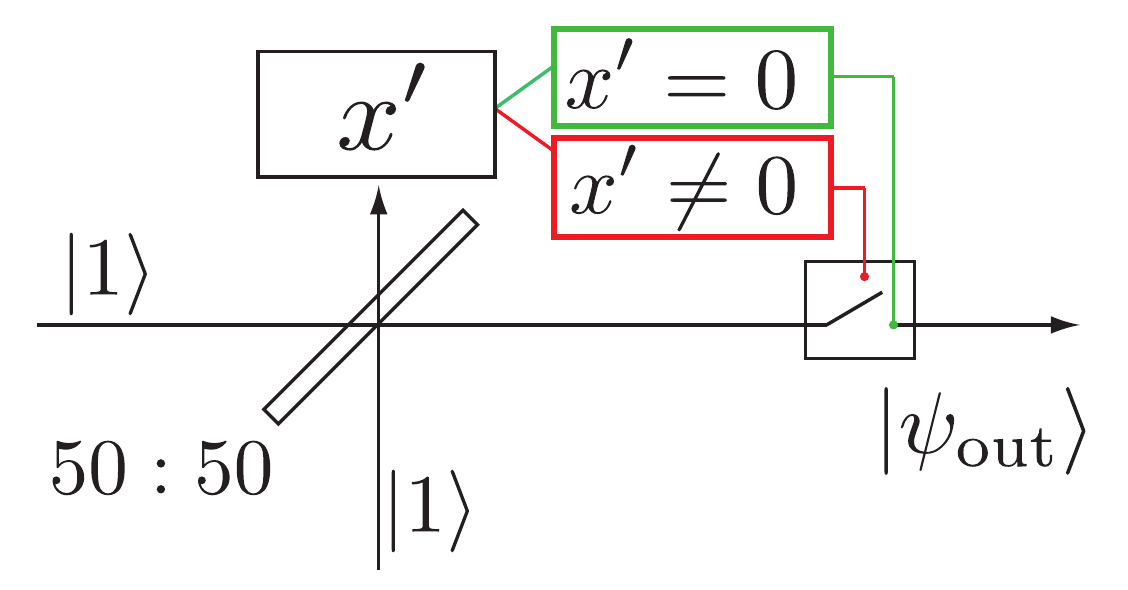}
\caption{Setup for the generation of an even squeezed cat state}
\label{principe}
\end{center}
\end{figure}

Optical Schr\"odinger cat states (SCS) of light, consisting in a coherent superposition of two coherent states, are of great interest in quantum optics as they have potential applications in various domains such as continuous-variable quantum computation \cite{Neergaard-Nielsen10,Ralph03}, quantum error-correcting codes \cite{Cochrane99, Gottesman01}, fundamental testings \cite{Sanders92,Wenger03,Jeong03,Stobinska07,Jeong08,Etesse14} or precision measurement \cite{Ralph02,Munro02,Joo11}. The main challenge in almost all these applications is to generate states whose amplitude is large enough to perform good quality operations \cite{Ralph03,Lund08}.\\
Usual strategies for the generation of free propagating SCS consist in heralding their preparation, either by click counting \cite{Ourjoum06,Gerrits10,Yukawa13} or by homodyne conditioning \cite{Ourjoum07}. But the main drawback of these techniques is that their success probability drops quickly with an increasing size of the output state. A first possibility to overcome this issue is to work with cubic non-linearities in cavities \cite{Yurke86,Turchette95,Jeong04,He09,Marek11,Vlastakis13}, but the trapped state cannot be used for quantum communication protocols. Another possibility is to increase the size of the SCS iteratively: mixing two SCS with small amplitude on a beamsplitter produces bigger SCS if the proper measurement is performed on one output arm. This ``breeding" operation can be done by click counting \cite{Lund04,Suzuki06}, but the success probability inherent to this kind of detector is too low to incorporate it in a realistic iterated protocol.\\
Homodyne heralding \cite{Laghaout13,Etesse14} seems to be a better candidate for SCS breeding operation, as the detection efficiency has no impact on the success probability of the whole operation. An important advantage of this kind of iterated methods is that between each iteration, a quantum memory can be incorporated in order to increase the success probability of the whole process.\\
In the present paper we have performed what is, to our knowledge, the first experimental realization of this kind of protocol \cite{Etesse14}. Let us briefly remind its principle: two single photon Fock states $\ket{1}$ are sent on a symmetrical beamsplitter (called breeding beamsplitter in the following), and a homodyne measurement is performed in one of the two output arms, as shown on Fig. \ref{principe}.
 If this measurement is equal to zero, the other arm is projected on the state:
\begin{eqnarray}
\psi_{\rm out}(x)&\propto&\varphi_2(0)\ket{0}-\varphi_0(0)\ket{2}\nonumber\\
&=&\frac{1}{\sqrt{3}}\ket{0}+\sqrt{\frac{2}{3}}\ket{2},
\label{etatcree}
\end{eqnarray}
where $\varphi_i(x)$ is the wavefunction of the Fock state $|i\rangle$.\\
This state is very close (99\% fidelity) to an even SCS of amplitude $\alpha=1.63$ squeezed by $s=1.52$ along the quadrature $x$:
\begin{equation}
\psi_{\rm cat}(x)\simeq 0.61\ket{0}+0.79\ket{2}+...,
\label{chatparf}
\end{equation}
where the following terms are non zero but can be neglected.\\
\begin{figure*}
\begin{center}
\subfigure[]{
\includegraphics[width=11cm]{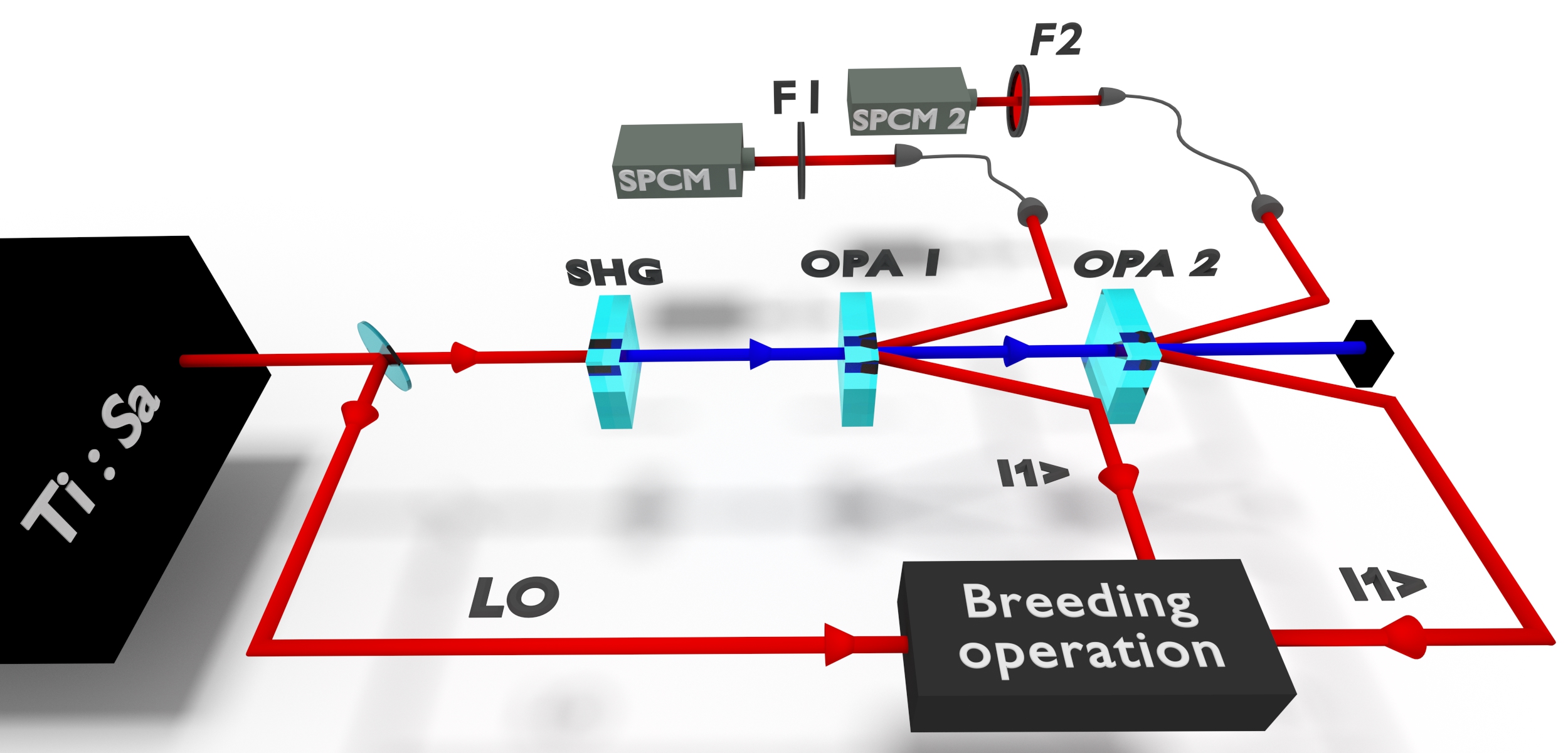}}
\subfigure[]{
\includegraphics[width=6.2cm]{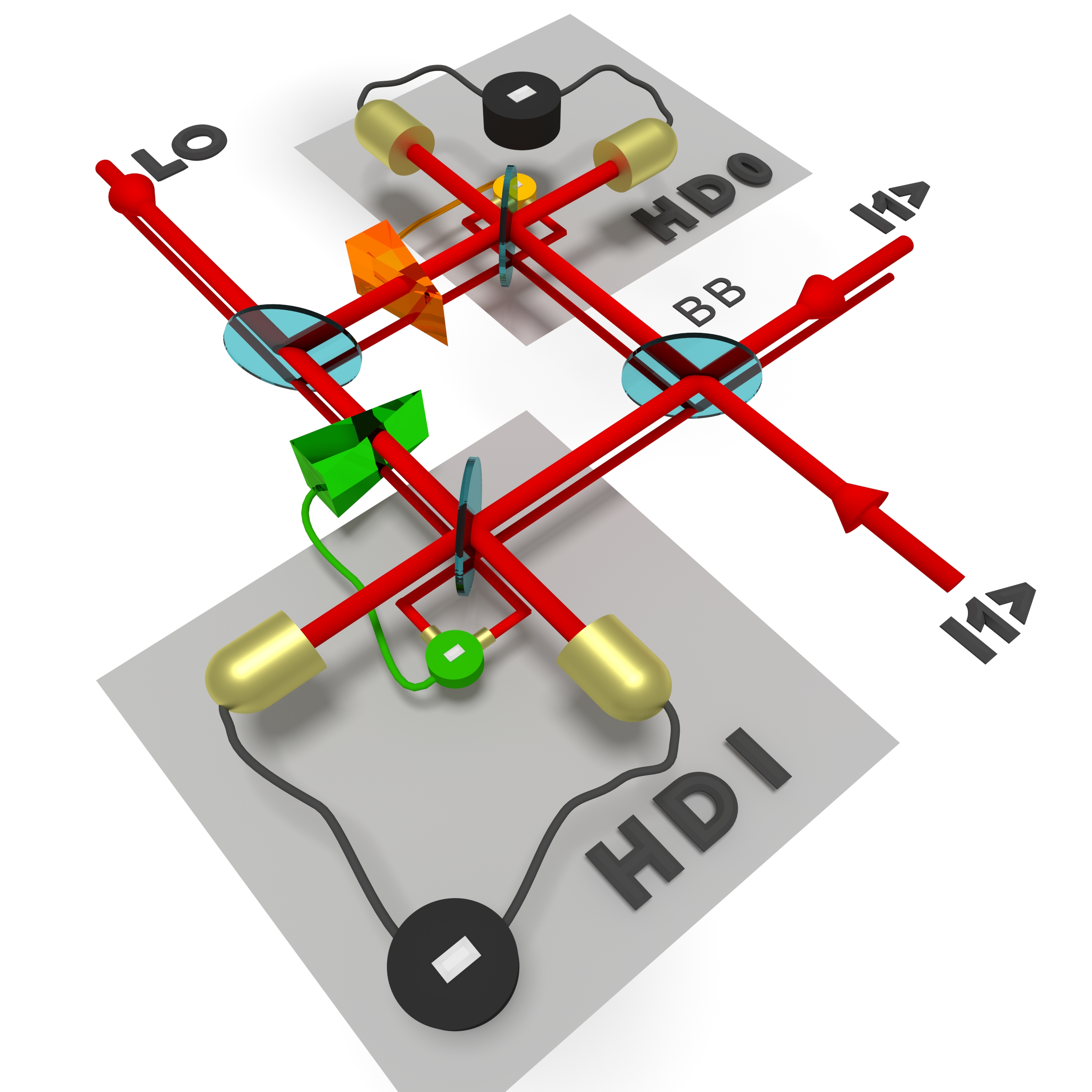}}
\caption{Experimental setup for the generation of squeezed Schr\"odinger cat state. (a) Setup for the generation of the two single photon Fock states $\ket{1}$ and the local oscillator (LO). SHG: second harmonic generation, OPA: optical parametric amplifier, F: spectral filter, SPCM: single photon counting module. (b) The breeding operation involved in the left figure. HD: homodyne detection, BB: breeding beamsplitter. The phase stabilization is performed with piezo mounted mirrors, symbolized by prisms of variable insertion in the figure.}
\label{expsetup}
\end{center}
\end{figure*}
Of course, conditioning on the exact measurement $x'=0$ will lead to a zero success probability, so experimentally one has to accept events within a window $x'\in[-\Delta x^c,\Delta x^c]$. Playing on $\Delta x^c$ will impact as well the success probability (improvement with increasing $\Delta x^c$) as the fidelity of the output state (improvement with decreasing $\Delta x^c$).\\
The experimental setup is presented on Fig. \ref{expsetup}. We use a Ti:Sa pulsed laser at 850 nm central wavelength, equipped with a cavity dumper which extracts 2 ps pulses of 40 nJ energy at a 800 kHz repetition rate. After stabilization and proper spatial shaping, the beam is focused in a 1.5 mm thick a-cut KNbO${}_3$ crystal in order to produce the necessary blue pump at 425 nm by second harmonic generation (SHG). The 14 nJ pulses (45\% conversion efficiency) of blue are then loosely focused (150 $\mu$m beam waist) inside two successive identical 1 mm thick a-cut KNbO${}_3$ crystals, in order to produce pairs of photons by spontaneous parametric down-conversion (SPDC). The two parametric gains $g_1=1.03$ and $g_2=1.01$ are sufficiently low to ensure that the generation of double pairs is rare enough, and that we have at most one photon in both arms in the proper mode. The SPDC occurs in a frequency degenerate but spatially nondegenerate configuration, allowing for a simple separation of the signal and idler beams. The two idler beams are coupled through two different 2-meters polarization maintaining single mode fibers towards single photon counting module (SPCM, \textit{Perkin Elmer SPCM-AQR-13}). A spectral filtering is performed on each beam with a grating and a slit at the focus of a lens (denoted by F on Fig \ref{expsetup}(a)). A click on the SPCM finally projects the signal beam into a single photon Fock state, whose quality depends on overall losses, generation imperfections and filtering quality. A model of imperfections can be found in \cite{Ourjoum06,Tualle-Brouri09,Ourjoum06b}: if we consider that the parametric amplification occurs with a gain $g$, and an excess noise gain $h$, the Wigner function of the imperfect single photon Fock state can be written as:
\begin{equation}
W_1(x,p)=\frac{e^{-\frac{x^2+p^2}{\sigma^2}}}{\pi\sigma^2}\Big[1-\delta+\frac{\delta (x^2+p^2)}{\sigma^2}\Big],
\label{eqWigPhotun}
\end{equation}
with
\begin{equation}
\sigma^2=2\eta(hg-1)+1\qquad
\delta=\frac{2\xi\eta h^2g(g-1)}{\sigma^2(hg-1)},
\end{equation}
and with $\eta$ the losses in the path of the state. A modal purity $\xi$ has also been introduced to account for the clicks on the SPCM that don't lead to a photon in the proper mode on the other arm. The parameter $\delta$ characterizes the quality of the state, as it reveals the negativity of the Wigner function if $\delta>1$ (cf Eq. (\ref{eqWigPhotun})).\\
The experimentally produced two single photon Fock states, which could be fairly assumed to be identical, are then temporally synchronized and mixed together on a 50:50 beamsplitter (the breeding beamsplitter) in order to get the Hong Ou Mandel effect \cite{Hong87}. The two-mode state is measured with two different Homodyne Detections (HD~0 and HD 1 on Fig. \ref{expsetup}(b)): one is used for the conditioning (HD 0) and the other one is used for the quantum state tomography of the conditioned output state (HD 1).
As the counting rate of the single photons is of the order of 1600 s${}^{-1}$, the count rate of simultaneous clicking of the two SPCM reduces to 3 s${}^{-1}$. The homodyne conditioning is chosen equal to $\Delta x^c=0.2$ (with a variance of vacuum equal to 1), which leads to a selection of approximately 15\% of the data. This last conditioning lowers the counting rate down to 27 min${}^{-1}$.
\begin{figure*}
\begin{center}
\includegraphics[width=18cm]{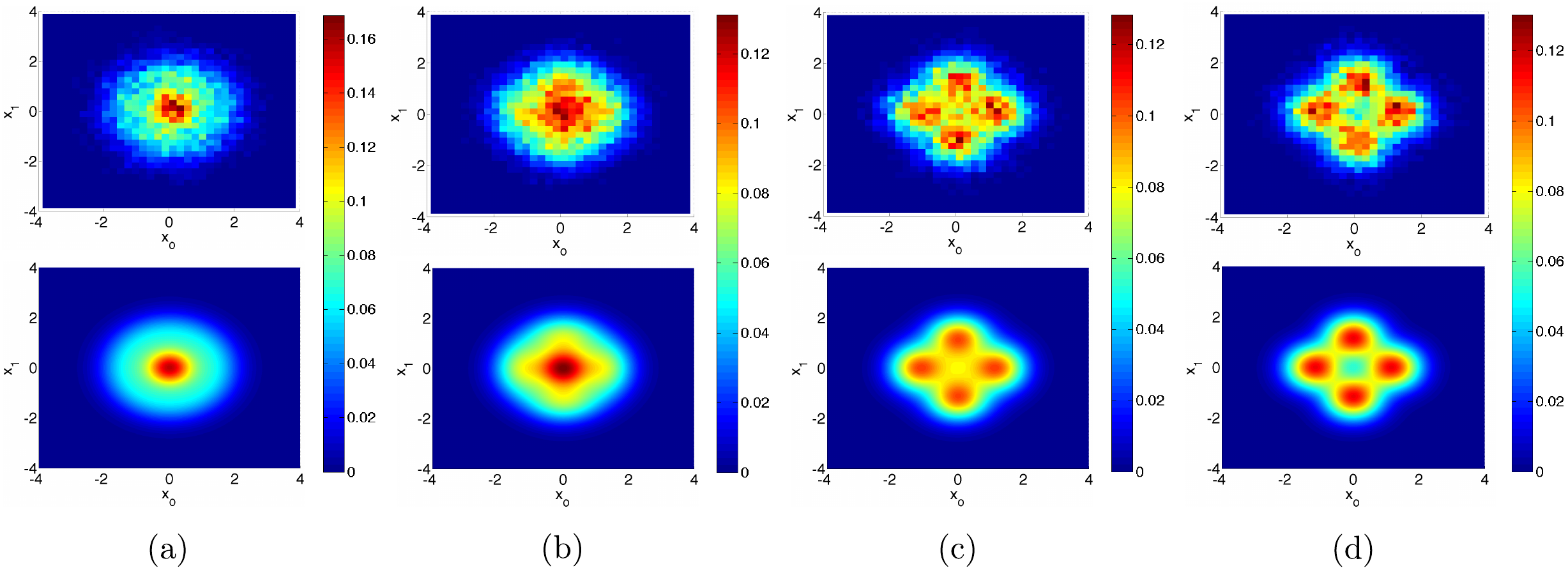}
\caption{Top row: 2D histograms of the experimental distributions of quadrature points. Bottom row: theoretical probability distributions of the homodyne points. Both are plotted for a relative phase between the two homodyne detections of (a) 90${}^{\circ}$, (b) 120${}^{\circ}$, (c) 150${}^{\circ}$ and (d) 180${}^{\circ}$. $x_0$: points measured by HD0, $x_1$: points measured by HD1.}
\label{histo2D}
\end{center}
\end{figure*}
Although the single photon Fock states are phase-invariant, the output state (\ref{etatcree}) clearly displays a phase dependence. This imposes to find a way to lock the relative phases of the two homodyne detections. A simple method of performing it is to superimpose the two paths by encoding them on orthogonal polarizations \cite{Ourjoum09} (one polarization per homodyne detection) but this method requires the use of polarizing beamsplitters, which have usually poor loss performances ($\sim 2-3\%$ losses). At the cost of a more sophisticated setup, we have preferred to use non polarizing beamsplitters ($\leq1\%$ losses) in order to preserve the quality of the state as long as possible.

The technique to lock the phases of the two homodyne detections is presented on Fig. \ref{expsetup}(b): an ancillary beam propagating 6 mm below the main beam (thin red beam on the figure) is used. This beam is separated from the main path by simply using shifted mirrors. The difference signal between the two arms of each homodyne detection (orange circuit for HD 0, green circuit for HD 1) gives the signal to stabilize, what is achieved by a feedback loop using fast piezo mounted mirrors driven with proportional integral circuits (one per homodyne detection). The phase of the main beam is then sequentially estimated by sending a coherent beam on the rear of a high reflectivity mirror, to get a weak coherent beam ($\sim$ 10 photons in average) entering the homodyne detections instead of the single photon state. A phase stability of $\pm 7^{\circ}$ has been achieved with this whole setup.\\

\begin{figure}[!h]
\begin{center}
\includegraphics[width=8.6cm]{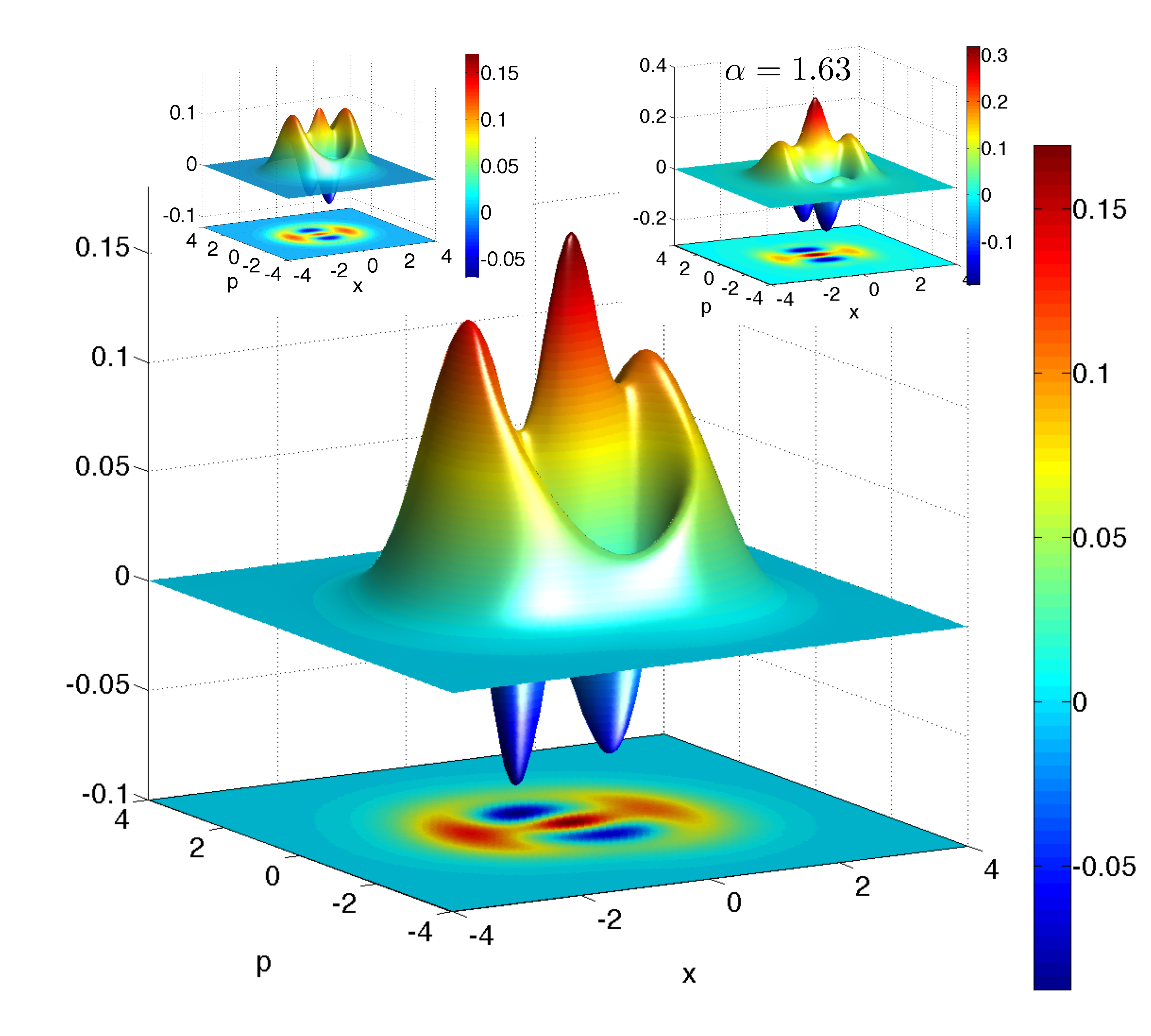}
\caption{Reconstructed Wigner function of the produced state, with 77\% detection efficiency correction. Inset left: Wigner function (\ref{formuleWignerOut}) obtained with the model (\ref{eqWigPhotun}) of imperfections ($\sigma=1.02$ and $\delta=1.17$). Inset right: Wigner function of the expected state (squeezed SCS of amplitude $\alpha=1.63$ and squeezed by a factor $s=1.52$)}
\label{WignerReconstr}
\end{center}
\end{figure}

Using this technique, we have acquired $\sim$ 15000 points (without homodyne heralding at this stage) for each of the four relative angles 90${}^{\circ}$, 120${}^{\circ}$, 150${}^{\circ}$ and 180${}^{\circ}$. As we will see later, due to symmetry reasons, these angles are sufficient to reconstruct the Wigner function. The 2D-histograms are shown on the first row of Fig. \ref{histo2D}, in which the data of the HD 1 are represented as a function of the data of the HD 0. As the state is entangled, the shape of these histograms cannot be factorized in the form $P(x_0,x_1)=P(x_0)P(x_1)$ for all the quadratures.\\
Given the Wigner function of the experimental single photon Fock state (\ref{eqWigPhotun}), we can calculate the theoretical probability distribution of the homodyne points according to the formula
  \begin{eqnarray}
 &P(x_0,x_{1})=\nonumber\\
 &\int\int  W_1\Big(\frac{x_0+X}{\sqrt{2}},\frac{p_0+P}{\sqrt{2}}\Big)W_1\Big(\frac{x_0-X}{\sqrt{2}},\frac{p_0-P}{\sqrt{2}}\Big)dp_0dp_{1},
 \end{eqnarray}
 with $X=x_{1}\cos(\theta)-p_{1}\sin(\theta)$, $P=p_{1}\cos(\theta)+x_{1}\sin(\theta)$ and $\theta$ the relative phase between the two homodyne detections. After calculation, we find:
\begin{eqnarray}
&P(x_0,x_{1})=\frac{e^{-\frac{x_0^2+{x^2_{1}}}{\sigma^2}}}{4\pi\sigma^6}\Big[\delta^2\big(x_0^2+{x^2_{1}}\big)^2\nonumber\\
&+\big(x_0^2+{x^2_{1}}\big)\sigma^2\delta(2\delta\cos^2\theta+4-4\delta)-4\delta^2x_0^2{x^2_{1}}\cos^2\theta\nonumber\\
&+\sigma^4\big(4-4\delta+\delta^2(2-\cos^2\theta)\big)\Big].
\label{distribBimod}
\end{eqnarray}
The angular dependence being in $\cos^2(\theta)$, this form confirms that the measurement of the quadratures of phase varying between 90${}^{\circ}$ and 180${}^{\circ}$ are sufficient to reconstruct the whole probability distribution.
By fitting our data with this theoretical distribution, we can recover the parameters $\delta$ and $\sigma$ of the single photon Fock state. The bottom row of figure \ref{histo2D} shows this theoretical distributions for the parameters $\sigma = 1.02$ and $\delta=1.17$, revealing a good agreement with the experimental data. As we have seen, the parameter $\delta$ is of great interest in the diagnosis of the quality of the input photons. Estimating it quickly is then central for an experimental realization, but the acquisition of the distributions involved in Eq. (\ref{distribBimod}) is too slow for a real-time diagnosis. We have then used another method, consisting in sending only one single photon in the protocol, and in performing the transformation $x'_1=\frac{x_0-x_1}{\sqrt{2}}$, which virtually removes the beamsplitter. By using the same method as in \cite{Ourjoum06b} it is then possible to estimate quickly the parameter $\delta$. With this measurement, the operator can decide whether the experiment can be performed or if it requires additional adjustments.\\
Eventually, in order to herald the generation of the expected squeezed SCS, we have to perform the homodyne conditioning on the HD 0 by selecting a vertical zone in the data presented in the top row of Fig. \ref{histo2D}: $x=x'\in[-\Delta x^c,\Delta x^c]$. It should be noticed that even if in our case we post-process the data, the conditioning can be performed in real time. By choosing $\Delta x^c=0.2$ as previously mentioned, we get $\sim$ 2000 homodyne conditioned points per phase. By using the Maximum Likelihood Estimation technique \cite{Lvovsky04} we could retrieve the Wigner function with these points, by taking into account 77\% of homodyne efficiency (0.946 photodiodes efficiency, $0.91^2$ local oscillator/signal matching and $0.99$ transmission). This function is shown on Fig. \ref{WignerReconstr}, and clearly displays two negative parts. The fidelity of this state with the expected squeezed SCS of amplitude $\alpha= 1.63$ and squeezed by $s=1.52$ along the $x$ quadrature is 61\%, which makes our state the highest amplitude and fidelity freely propagating even SCS ever produced. Given our analytical model, it is also possible to reconstruct the Wigner function of the state. In order to simplify the calculation, one can assume a conditioning performed according to a Gaussian law rather than a square law. Such an approximation has only small influence on the shape of the Wigner function, which can then be simply written as:
  \begin{eqnarray}
 &W(x_1,p_{1})=\label{formuleWignerOut}\\
 &\int\int  W_1\Big(\frac{x_0+x_1}{\sqrt{2}},\frac{p_0+p_1}{\sqrt{2}}\Big)W_1\Big(\frac{x_0-x_1}{\sqrt{2}},\frac{p_0-p_1}{\sqrt{2}}\Big)e^{-\frac{x^2_0}{2(\Delta {x^c})^2}}dx_1dp_{1}.\nonumber
 \end{eqnarray}
For comparison, the Wigner function (\ref{formuleWignerOut}) is plotted in the left inset of Fig. \ref{WignerReconstr}, showing great similarity with the previously reconstructed function (94\% fidelity \footnote{Here we use the definition of fidelity between two states $\rho_1$ and $\rho_2$: $F=\text{Tr}^2(\sqrt{\sqrt{\rho_1}\rho_2\sqrt{\rho_1}})$}).\\
Given the fact that the number of samples per phase is quite low ($\sim2000$), we have also performed statistical error estimation of our results. To achieve this, a Monte Carlo simulation based on the density matrix found by maximum likelihood estimation has been performed \cite{Lvovsky04}. With this error estimation, the negativity of the Wigner function is -0.08$\pm$0.01 with detection efficiency correction and -0.024$\pm$0.01 without this correction, revealing the strong non-classical feature of our state even without post-processing. The fidelity itself is also subject to a statistical uncertainty, which is of the order of 61$\pm1 \%$ with correction in our case, showing the good quality of the created state.

A very interesting point is that this experimental realization is the first stage of a protocol of broader interest \cite{Etesse14}: by iterating the operation described in the present paper, \textit{i. e.} by feeding the protocol with the states we created instead of the single photons, it is possible to grow the size of the cat state. In other words, we have presented a proof of concept for the cornerstone of cat ``breeding" operation.\\
In conclusion, we have experimentally realized the generation of an even squeezed SCS by the use of a new method based on homodyne conditioning and requiring single photon Fock states in input only. The fidelity of the created state is 61\% with a SCS of amplitude $\alpha=1.63$ and squeezed by a factor of $s=1.52$ along the quadrature $x$. Due to the structure of the protocol, it is possible to cascade it in order to grow the size of the cat states iteratively.
With the help of quantum memories, which are currently in rapid development \cite{Yoshikawa13}, we believe in the strong interest of this iterative method compared to the commonly used photon-subtraction methods.

\begin{acknowledgments}
We acknowledge support from the labex PALM project HAQI, and from the French National Research Agency Project SPOCQ.
\end{acknowledgments}

\bibliography{Paper251114}
\end{document}